\documentclass[english,english,english]{article}
\usepackage[T1]{fontenc}
\usepackage[latin1]{inputenc}
\usepackage{a4wide}
\usepackage{babel}
\makeatletter

\providecommand{\LyX}{L\kern-.1667em\lower.25em\hbox{Y}\kern-.125emX\@}

\usepackage[T1]{fontenc}
\usepackage[latin1]{inputenc}
\usepackage{a4wide}
\usepackage{babel}
\makeatletter

\usepackage[T1]{fontenc}
\usepackage[latin1]{inputenc}
\usepackage{a4wide}
\usepackage{babel}


\makeatother
\begin{document}
{\large \hfill{}}\textbf{\large On local indistinguishability of orthogonal
pure states by using a bound\hfill{} }{\large \par}

\textbf{\large \hfill{}on distillable entanglement}{\large \hfill{} }{\large \par}

\hfill{}Sibasish Ghosh\( ^{a} \)\footnote{%
sghosh@starsky.ee.ucla.edu 
}, Guruprasad Kar\( ^{b} \)\footnote{%
gkar@isical.ac.in 
}, Anirban Roy\( ^{b} \)\footnote{%
res9708@isical.ac.in 
}, Debasis Sarkar\( ^{c} \)\footnote{%
dsarkar@cubmb.ernet.in 
}, Aditi Sen(De)\( ^{d} \)\footnote{%
aditi@iftia6.univ.gda.pl 
} \hfill{} 

\hfill{}and Ujjwal Sen\( ^{e} \)\footnote{%
ujjwal@isiosf.isi.it 
}\hfill{} 

{\footnotesize \hfill{}\( ^{a} \)}\emph{\footnotesize Electrical
Engineering Department, University of California, Los Angeles, Los
Angeles, CA 90095, USA}{\footnotesize \hfill{}} {\footnotesize \par{}}{\footnotesize \par}

{\footnotesize \hfill{}\( ^{b} \)}\emph{\footnotesize Physics and
Applied Mathematics Unit, Indian Statistical Institute, 203 BT Road,
Kolkata 700035, India}{\footnotesize \hfill{}} {\footnotesize \par{}}{\footnotesize \par}

{\footnotesize \hfill{}\( ^{c} \)}\emph{\footnotesize Department
of Applied Mathematics, University of Calcutta, 92 APC Road, Kolkata
700009, India}{\footnotesize \hfill{}} {\footnotesize \par{}}{\footnotesize \par}

{\footnotesize \hfill{}\( ^{d} \)}\emph{\footnotesize Institute of
Theoretical Physics and Astrophysics, University of Gda\'{n}sk, 80-952
Gda\'{n}sk, Poland}{\footnotesize \hfill{}} {\footnotesize \par{}}{\footnotesize \par}

{\footnotesize \hfill{}\( ^{e} \)}\emph{\footnotesize Institute for
Scientific Interchange (ISI) Foundation, Viale Settimio Severo 65,
I-10133 Torino, Italy}{\footnotesize \hfill{}} {\footnotesize \par{}}{\footnotesize \par}

\begin{abstract}
We show that the four states \( a\left| 00\right\rangle +b\left| 11\right\rangle  \),
\( \overline{b}\left| 00\right\rangle -\overline{a}\left| 11\right\rangle  \),
\( c\left| 01\right\rangle +d\left| 10\right\rangle  \) and \( \overline{d}\left| 01\right\rangle -\overline{c}\left| 10\right\rangle  \)
cannot be discriminated with certainty if only local operations and
classical communication (LOCC) are allowed and if only a single copy
is provided, except in the case when they are simply \( \left| 00\right\rangle  \),
\( \left| 11\right\rangle  \), \( \left| 01\right\rangle  \) and
\( \left| 10\right\rangle  \) (in which case they are trivially distinguishable
with LOCC). We go on to show that there exists a continuous range
of values of \( a \), \( b \), \( c \) and \( d \) such that even
three states among the above four are not locally distinguishable,
if only a single copy is provided. The proof follows from the fact
that logarithmic negativity is an upper bound of distillable entanglement.
\end{abstract}
Entanglement \cite{key-15} has always been a storehouse of surprises.
It has been the vehicle in demonstration of several paradoxes \cite{key-16, key-12, key-13, key-14}.
However in the past few years, entanglement has been found to be useful
in information processing and communication between possibly distant
parties which \emph{a priori} share an entangled state \cite{key-0}.
Examples include quantum cryptography \cite{key-21}, dense coding
\cite{key-19}, quantum teleportation \cite{key-17}, enhanced communication
\cite{key-20}. As the focus is on what can and cannot be implemented
between separated parties when they \emph{a priori} share some entangled
state, greater attention is being given on what can and cannot be
done with entangled states, when we act locally on them. For example,
it has recently been shown that given a single copy from a set of
any \emph{two} multipartite orthogonal pure states, it is always possible
to distinguish between them even if one is acting only locally \cite{key-1}.
It is therefore natural to probe the question of local distinguishability
of a set containing more than two orthogonal (in general, entangled)
states. Our attempt in this paper is to investigate the question of
local indistinguishability of bipartite (\( 2\otimes 2 \)) states
which is a recent interest in understanding entanglement \cite{key-1, key-2, key-3, key-4}. 

In general, more than two orthogonal states cannot be discriminated.
For example, any three of the four Bell states \begin{equation}
\label{bell}
\begin{array}{rcl}
\displaystyle \left| B_{1}\right\rangle  & = & {\frac{1}{\sqrt{2}}\left( \left| 00\right\rangle +\left| 11\right\rangle \right) },\\
\left| B_{2}\right\rangle  & = & {\frac{1}{\sqrt{2}}\left( \left| 00\right\rangle -\left| 11\right\rangle \right) },\\
\left| B_{3}\right\rangle  & = & {\frac{1}{\sqrt{2}}\left( \left| 01\right\rangle +\left| 10\right\rangle \right) },\\
\left| B_{4}\right\rangle  & = & {\frac{1}{\sqrt{2}}\left( \left| 01\right\rangle -\left| 10\right\rangle \right) },
\end{array}
\end{equation}
 cannot be deterministically discriminated by using local operations
and classical communication (LOCC) when only a single copy is provided
\cite{key-4}. However the problem of discrimination of an arbitrary
set of three or four orthogonal (in general, entangled) states in
\( 2\otimes 2 \) seems to be quite formidable.

In this paper we probe the question of local distinguishability of
the following set of four (orthogonal) states: \begin{equation}
\label{nonbell}
\begin{array}{rcl}
\displaystyle \left| A_{1}\right\rangle  & = & {a\left| 00\right\rangle +b\left| 11\right\rangle },\\
\left| A_{2}\right\rangle  & = & {\overline{b}\left| 00\right\rangle -\overline{a}\left| 11\right\rangle },\\
\left| A_{3}\right\rangle  & = & {c\left| 01\right\rangle +d\left| 10\right\rangle },\\
\left| A_{4}\right\rangle  & = & {\overline{d}\left| 01\right\rangle -\overline{c}\left| 10\right\rangle }
\end{array}
\end{equation}
 We show that these four states cannot be discriminated deterministically
by using LOCC when only a single copy is provided, except when the
\( \left| A_{i}\right\rangle  \)'s are just \( \left| 00\right\rangle  \),
\( \left| 11\right\rangle  \), \( \left| 01\right\rangle  \) and
\( \left| 10\right\rangle  \) (in which case they are trivially distinguishable
with LOCC) \cite{key-8}. We are therefore faced with the question
as to whether any \emph{three} among the above four states in (2)
are locally distinguishable. We show that for a certain continuous
range of values of \( a \), \( b \), \( c \) and \( d \), even
\emph{three} of the above four states in (2) cannot be deterministically
discriminated if only LOCC are allowed and if only a single copy is
provided. However this continuous range of values of \( a \), \( b \),
\( c \) and \( d \) do \emph{not} include values to reproduce the
set of four Bell states from the set of \( \left| A_{i}\right\rangle  \)'s
and hence our results do not reproduce the result obtained in \cite{key-4}.
Without loosing any generality, we assume here that \( |a|\geq |b| \)
and \( |c|\geq |d| \).

We first prove that the four states \( \left\{ \left| A_{i}\right\rangle \right\}  \)
in (\ref{nonbell}) cannot be discriminated with certainty if only
LOCC are allowed and if only a single copy is provided, except when
the \( \left| A_{i}\right\rangle  \)'s are \( \left| 00\right\rangle  \),
\( \left| 11\right\rangle  \), \( \left| 01\right\rangle  \) and
\( \left| 10\right\rangle  \). To prove it, we exploit a property
of a function called logarithmic negativity (\( E_{N}(\rho ) \))
\cite{key-5} of the state parameters of a bipartite state \( \rho  \).
It is defined as \( E_{N}(\rho )\equiv {\textrm{log}}_{2}\left\Vert \rho ^{T_{A}}\right\Vert _{1} \)
for a state \( \rho _{AB} \) of two parties A and B. The trace norm
of a square matrix \( \sigma  \) is denoted by \( \left\Vert \sigma \right\Vert _{1} \),
and defined as \( \left\Vert \sigma \right\Vert _{1}\equiv {\textrm{Tr}}[{(\sigma ^{\dagger }\sigma )^{1/2}}] \).
And here \( \rho ^{T_{A}} \) is the partial transpose \cite{key-23}
of \( \rho _{AB} \) with respect to the part A of the bipartite state
\( \rho _{AB} \). It turns out that we can express \( E_{N}(\rho ) \)
as \( E_{N}(\rho )=\log _{2}\left( 1+2N(\rho )\right)  \), where
\( N(\rho ) \) is the absolute value of the sum of the negative eigenvalues
of \( \rho ^{T_{A}} \). The property of logarithmic negativity that
we use here is that it is an upper bound of distillable entanglement
\cite{key-5}. This property has recently been used in demonstration
of irreversibility in asymptotic manipulations of entanglement \cite{key-24}.

Consider the following state shared between Alice (A), Bob (B), Charu
(C) and Debu (D), with all four at distant locations:\[
\rho =\frac{1}{4}\sum ^{4}_{i=1}P\left[ \left| A_{i}\right\rangle _{AB}\left| B_{i}\right\rangle _{CD}\right] \]
 Here the \( \left| A_{i}\right\rangle  \)'s are given by equation
(\ref{nonbell}) and \( \left| B_{i}\right\rangle  \)'s are given
by equation (\ref{bell}) \cite{key-22}. Suppose that it is possible
to distinguish the four states \( \left\{ \left| A_{i}\right\rangle \right\}  \)
with certainty even if only LOCC are allowed and only a single copy
is provided. Then it immediately follows from the structure of the
shared state \( \rho  \) that Alice and Bob (without meeting) would
be able to help Charu and Debu to share a Bell state with certainty.

This means that the distillable entanglement of \( \rho  \), in the
AC:BD cut, is at least \( 1 \) ebit.

Now the logarithmic negativity \( E_{N}(\rho )\equiv {\textrm{log}}_{2}\left\Vert \rho ^{T_{AC}}\right\Vert _{1} \)
of the state \( \rho  \), in the AC:BD cut, is \[
\log _{2}\left( \left| a\right| ^{2}+\left| c\right| ^{2}\right) ,\]
 which is strictly less than unity, except when the \( \left| A_{i}\right\rangle  \)'s
are \( \left| 00\right\rangle  \), \( \left| 11\right\rangle  \),
\( \left| 01\right\rangle  \) and \( \left| 10\right\rangle  \)
\cite{key-10}.

However, \( E_{N}(\rho ) \) is an upper bound of distillable entanglement
\cite{key-5}. This implies that the distillable entanglement of the
state \( \rho  \) in the AC:BD cut must be strictly less than 1 ebit.
But as we have already stated, the assumption of local distinguishability
of the \( \left| A_{i}\right\rangle  \)'s forces the distillable
entanglement of \( \rho  \) in the AC:BD cut to be at least 1 ebit.
This is a contradiction. Thus our assumption on the local distinguishability
of the \( \left| A_{i}\right\rangle  \)'s is proved to be wrong.
In other words, we have proved that the four (orthogonal) states \( \left\{ \left| A_{i}\right\rangle \right\}  \)
cannot be distinguished locally, with certainty, if only a single
copy is provided (except in the trivial case when the states are \( \left| 00\right\rangle  \),
\( \left| 11\right\rangle  \), \( \left| 01\right\rangle  \) and
\( \left| 10\right\rangle  \)).

We now go on to prove that there exists a certain continuous range
of values of \( a \), \( b \), \( c \) and \( d \), for which
three of the four states from the set \( \left\{ \left| A_{i}\right\rangle \right\}  \)
in equation (\ref{nonbell}) cannot be deterministically discriminated
if only LOCC are allowed and if only a single copy is provided.

Consider the following state shared between Alice, Bob, Charu and
Debu, with all four at distant locations: \[
\eta =\frac{1}{3}\sum ^{3}_{i=1}P\left[ \left| A_{i}\right\rangle _{AB}\left| B_{i}\right\rangle _{CD}\right] \]
\( \left| A_{i}\right\rangle  \)'s are given by equation (\ref{nonbell})
and \( \left| B_{i}\right\rangle  \)'s are given by equation (\ref{bell})
\cite{key-11}. Again we suppose that it is possible to locally distinguish
with certainty, the three states \( a\left| 00\right\rangle +b\left| 11\right\rangle  \),
\( \overline{b}\left| 00\right\rangle -\overline{a}\left| 11\right\rangle  \)
and \( c\left| 01\right\rangle +d\left| 10\right\rangle  \), even
if only a single copy is provided. And again, as earlier, it implies
that the distillable entanglement of the state \( \eta  \) (in the
AC:BD cut) is more than or equal to 1 ebit.

The logarithmic negativity \cite{key-5} of \( \eta  \), in the AC:BD
cut is \[
\log _{2}\left\{ \frac{1}{3}\left( \sqrt{1+16\left| ab\right| ^{2}-4\left| cd\right| ^{2}}+2\sqrt{1-4\left| ab\right| ^{2}+\left| cd\right| ^{2}}\right) +1\right\} .\]
 There would again arise a contradiction, if this expression is strictly
less than unity.

This implies that whenever we have \begin{equation}
\label{cond}
4\left| ab\right| ^{2}-\left| cd\right| ^{2}>3/4,
\end{equation}
 the states \( a\left| 00\right\rangle +b\left| 11\right\rangle  \),
\( \overline{b}\left| 00\right\rangle -\overline{a}\left| 11\right\rangle  \)
and \( c\left| 01\right\rangle +d\left| 10\right\rangle  \) would
be locally indistinguishable with certainty, if only a single copy
is provided. It is obvious from the expression on the left hand side
of (\ref{cond}), that the condition would not change if \( c\left| 01\right\rangle +d\left| 10\right\rangle  \)
is replaced by \( \overline{d}\left| 01\right\rangle -\overline{c}\left| 10\right\rangle  \).
And the condition would change to \begin{equation}
\label{cond1}
4\left| cd\right| ^{2}-\left| ab\right| ^{2}>3/4,
\end{equation}
 if we investigate the local indistinguishability of \( c\left| 01\right\rangle +d\left| 10\right\rangle  \),
\( \overline{d}\left| 01\right\rangle -\overline{c}\left| 10\right\rangle  \)
and any one of \( a\left| 00\right\rangle +b\left| 11\right\rangle  \)
and \( \overline{b}\left| 00\right\rangle -\overline{a}\left| 11\right\rangle  \).

It is interesting to note that none of these inequalities satisfy
the values of \( a \), \( b \), \( c \), \( d \) such that the
Bell states can be obtained from the \( \left| A_{i}\right\rangle  \)'s.
Therefore the result that any three Bell states cannot be discriminated
with certainty if only LOCC are allowed and if only a single copy
is provided \cite{key-4}, is not reproduced by the results of this
paper. This fact is quite plausible, because we have considered logarithmic
negativity \cite{key-5} as an upper bound of distillable entanglement,
while in ref. \cite{key-4}, relative entropy of entanglement \cite{key-6}
was taken as an upper bound of distillable entanglement \cite{key-27}.

It is interesting to consider the following cases:

\textbf{\underbar{Case (1.1.a)}} From (\ref{cond}), it follows that
the three states \( (1/{\sqrt{2}})\left( \left| 00\right\rangle +\left| 11\right\rangle \right)  \),
\( (1/{\sqrt{2}})\left( \left| 00\right\rangle -\left| 11\right\rangle \right)  \)
and \( c\left| 01\right\rangle +d\left| 10\right\rangle  \) (or \( \overline{d}\left| 01\right\rangle -\overline{c}\left| 10\right\rangle  \))
are locally indistinguishable with certainty if only a single copy
is provided, for \emph{all} values of \( c \) and \( d \) except
when \( \left| cd\right| =1/2 \).

\noindent In particular, the states \( \frac{1}{\sqrt{2}}\left( \left| 00\right\rangle +\left| 11\right\rangle \right)  \),
\( \frac{1}{\sqrt{2}}\left( \left| 00\right\rangle -\left| 11\right\rangle \right)  \)
and \( \left| 01\right\rangle  \) (or \( \left| 10\right\rangle  \))
are locally indistinguishable with certainty if only a single copy
is provided.

\textbf{\underbar{Case (1.1.b)}} It was shown in \cite{key-4} that
the three states \( (1/{\sqrt{2}})\left( \left| 00\right\rangle +\left| 11\right\rangle \right)  \),
\( (1/{\sqrt{2}})\left( \left| 00\right\rangle -\left| 11\right\rangle \right)  \)
and \( c\left| 01\right\rangle +d\left| 10\right\rangle  \) (or \( \overline{d}\left| 01\right\rangle -\overline{c}\left| 10\right\rangle  \))
are locally indistinguishable with certainty if only a single copy
is provided, for \emph{all} values of \( c \) and \( d \) when \( \left| cd\right| =1/2 \)
\cite{key-7}.

\noindent Combining the cases (1.1.a) and (1.1.b), we have the following
result:

\textbf{\underbar{Case (1.1)}} The three states \( (1/{\sqrt{2}})\left( \left| 00\right\rangle +\left| 11\right\rangle \right)  \),
\( (1/{\sqrt{2}})\left( \left| 00\right\rangle -\left| 11\right\rangle \right)  \)
and \( c\left| 01\right\rangle +d\left| 10\right\rangle  \) (or \( \overline{d}\left| 01\right\rangle -\overline{c}\left| 10\right\rangle  \))
are locally indistinguishable with certainty if only a single copy
is provided, for \emph{all} values of \( c \) and \( d \).

\noindent Similarly we have the following case, which follows from
equation (\ref{cond1}) and ref. \cite{key-4}:

\textbf{\underbar{Case (1.2)}} The three states \( (1/{\sqrt{2}})\left( \left| 01\right\rangle +\left| 10\right\rangle \right)  \),
\( (1/{\sqrt{2}})\left( \left| 01\right\rangle -\left| 10\right\rangle \right)  \)
and \( a\left| 00\right\rangle +b\left| 11\right\rangle  \) (or \( \overline{b}\left| 00\right\rangle -\overline{a}\left| 11\right\rangle  \))
are locally indistinguishable with certainty if only a single copy
is provided, for \emph{all} values of \( a \) and \( b \) \cite{key-7}.

\noindent On the other hand, one can see that

\textbf{\underbar{Case (2.1.a)}} the three states \( a\left| 00\right\rangle +b\left| 11\right\rangle  \),
\( \overline{b}\left| 00\right\rangle -\overline{a}\left| 11\right\rangle  \)
and \( (1/{\sqrt{2}})\left( \left| 01\right\rangle +\left| 10\right\rangle \right)  \)
(or \( (1/{\sqrt{2}})\left( \left| 01\right\rangle -\right.  \) \( \left. \left| 10\right\rangle \right)  \))
are (trivially) distinguishable with certainty by LOCC if \( ab=0 \)
even in the single copy case.

\noindent And (as in ref. \cite{key-4}) one can show that

\textbf{\underbar{Case (2.1.b)}} the three states \( a\left| 00\right\rangle +b\left| 11\right\rangle  \),
\( \overline{b}\left| 00\right\rangle -\overline{a}\left| 11\right\rangle  \)
and \( (1/{\sqrt{2}})\left( \left| 01\right\rangle +\left| 10\right\rangle \right)  \)
(or \( (1/{\sqrt{2}})\left( \left| 01\right\rangle -\right.  \) \( \left. \left| 10\right\rangle \right)  \))
are indistinguishable with certainty by LOCC if \( |ab|=1/2 \), in
the single copy case.

\noindent Similarly we have the following cases:

\textbf{\underbar{Case(2.2.a)}} The three states \( c\left| 01\right\rangle +d\left| 10\right\rangle  \),
\( \overline{d}\left| 01\right\rangle -\overline{c}\left| 10\right\rangle  \)
and \( (1/{\sqrt{2}})\left( \left| 00\right\rangle +\left| 11\right\rangle \right)  \)
(or \( (1/{\sqrt{2}})\left( \left| 00\right\rangle -\right.  \) \( \left. \left| 11\right\rangle \right)  \))
are (trivially) distinguishable with certainty by LOCC if \( cd=0 \)
even in the single copy case. 

\textbf{\underbar{Case(2.2.b)}} The three states \( c\left| 01\right\rangle +d\left| 10\right\rangle  \),
\( \overline{d}\left| 01\right\rangle -\overline{c}\left| 10\right\rangle  \)
and \( (1/{\sqrt{2}})\left( \left| 00\right\rangle +\left| 11\right\rangle \right)  \)
(or \( (1/{\sqrt{2}})\left( \left| 00\right\rangle -\right.  \) \( \left. \left| 11\right\rangle \right)  \))
are indistinguishable with certainty by LOCC if \( |cd|=1/2 \), in
the single copy case.

However the local distinguishability (with certainty and in the single
copy case) of the states \( a\left| 00\right\rangle +b\left| 11\right\rangle  \),
\( \overline{b}\left| 00\right\rangle -\overline{a}\left| 11\right\rangle  \)
and \( (1/{\sqrt{2}})\left( \left| 01\right\rangle +\left| 10\right\rangle \right)  \)
(or \( (1/{\sqrt{2}})\left( \left| 01\right\rangle -\left| 10\right\rangle \right)  \))
is still inconclusive for \emph{}all values of \( a \) and \( b \)
except in the cases when \( ab=0 \) or \( \left| ab\right| =1/2 \).
And similar is the situation for \( c\left| 01\right\rangle +d\left| 10\right\rangle  \),
\( \overline{d}\left| 01\right\rangle -\overline{c}\left| 10\right\rangle  \)
and \( (1/{\sqrt{2}})\left( \left| 00\right\rangle +\left| 11\right\rangle \right)  \)
(or \( (1/{\sqrt{2}})\left( \left| 00\right\rangle -\left| 11\right\rangle \right)  \))
for all values of \( c \) and \( d \) except when \( cd=0 \) or
\( \left| cd\right| =1/2 \).

In conclusion, we have shown that the four states \( a\left| 00\right\rangle +b\left| 11\right\rangle  \),
\( \overline{b}\left| 00\right\rangle -\overline{a}\left| 11\right\rangle  \),
\( c\left| 01\right\rangle +d\left| 10\right\rangle  \) and \( \overline{d}\left| 01\right\rangle -\overline{c}\left| 10\right\rangle  \)
cannot be distinguished with certainty if only local operations and
classical communication are allowed and if only a single copy is provided,
except when they are simply \( \left| 00\right\rangle  \), \( \left| 11\right\rangle  \),
\( \left| 01\right\rangle  \) and \( \left| 10\right\rangle  \)
(in which case they are trivially distinguishable with LOCC). We went
on to show that there exists a continuous range of values of \( a \),
\( b \), \( c \), \( d \) for which even \emph{three} of the above
four states are not locally distinguishable with certainty if only
a single copy is provided. Precisely, \( a\left| 00\right\rangle +b\left| 11\right\rangle  \),
\( \overline{b}\left| 00\right\rangle -\overline{a}\left| 11\right\rangle  \)
and \( c\left| 01\right\rangle +d\left| 10\right\rangle  \) (or \( \overline{d}\left| 01\right\rangle -\overline{c}\left| 10\right\rangle  \))
are not locally distinguishable with certainty, if only a single copy
is provided, when \( 4\left| ab\right| ^{2}-\left| cd\right| ^{2}>3/4 \). 

Let us add here that the relative entropy of entanglement \cite{key-6}
\( E_{R}(\eta ) \), of \( \eta =\frac{1}{3}\sum ^{3}_{i=1}P\left[ \left| A_{i}\right\rangle _{AB}\left| B_{i}\right\rangle _{CD}\right]  \)
in the AC:BD cut, is \( 2-\log _{2}3<0.42 \) \cite{key-28}, \emph{for}
\( \left| A_{i}\right\rangle =\left| B_{i}\right\rangle  \), with
\( \left| B_{i}\right\rangle  \)'s being any three Bell states. But
relative entropy of entanglement satisfies a strong continuity condition
\cite{key-25}. And if \( \left| A_{1}\right\rangle =a\left| 00\right\rangle +b\left| 11\right\rangle  \),
\( \left| A_{2}\right\rangle =\overline{b}\left| 00\right\rangle -\overline{a}\left| 11\right\rangle  \)
and \( \left| A_{3}\right\rangle =c\left| 01\right\rangle +d\left| 10\right\rangle  \),
then the \( \left| A_{i}\right\rangle  \)'s would be the Bell states
for \( (a,\: b,\: c,\: d)=\left( \frac{1}{\sqrt{2}},\: \frac{1}{\sqrt{2}},\: \frac{1}{\sqrt{2}},\: \frac{1}{\sqrt{2}}\right)  \),
which is a boundary point of the set of points represented by \( 4\left| ab\right| ^{2}-\left| cd\right| ^{2}\leq 3/4 \).
This implies, via the continuity of \( E_{R} \), that \( E_{R}(\eta )<1 \)
would hold at least for some continuous subset of the region \( 4\left| ab\right| ^{2}-\left| cd\right| ^{2}\leq 3/4 \).
And the fact that relative entropy of entanglement is an upper bound
of distillable entanglement \cite{key-27} would result in the corresponding
\( \left| A_{i}\right\rangle  \)'s being locally indistinguishable
(for a single copy) by the methodology used in this paper. However,
the value of \( E_{R}(\eta ) \) is not computable at present for
values of \( (a,\: b,\: c,\: d) \) lying the relevant range \cite{key-28}.
It is probably true that local indistinguishability with certainty
holds even when \( 4\left| ab\right| ^{2}-\left| cd\right| ^{2}\leq 3/4 \)
(leaving out the trivial case of \( \left| 00\right\rangle  \), \( \left| 11\right\rangle  \)
and \( \left| 01\right\rangle  \) (or \( \left| 10\right\rangle  \))).
But that would necessiate the consideration of a different upper bound
of distillable entanglement, or a different method than is followed
here.\\

The work of SG was sponsored in part by the Defense Advanced Research
Projects Agency (DARPA) project MDA 972-99-1-0017 (note that the content
of this paper does not necessarily reflect the position or the policy
of the government and no official endorsement should be inferred),
and in part by the U.S. Army Research Office/DARPA under contract/grant
number DAAD 19-00-1-0172. The work of AS is supported by the EU Project
EQUIP Contract No. IST-1999-11053. US acknowledges support by the
European Community through grant IST-1999-10596 (Q-ACTA).


\begin{thebibliography}{10}
\bibitem{key-8}{\small \( |0\rangle  \) and \( |1\rangle  \) are the eigen states
of the Pauli spin matrix \( \sigma _{z} \) corresponding to eigen
values \( 1 \) and \( -1 \). And the \( \left| A_{i}\right\rangle  \)'s
are normalised.}{\small \par}
\bibitem{key-10}{\small It is to be noted here that the quantity \( {\textrm{log}}_{2}\left( \left| a\right| ^{2}+\left| c\right| ^{2}\right)  \)
is non-negative, due to our assumptions that \( |a|\geq |b| \) and
\( |c|\geq |d| \).}{\small \par}
\bibitem{key-7}{\small Although in ref. \cite{key-4}, \( (c,\: d) \) is taken to
be either equal to \( (1/\sqrt{2},\: 1/\sqrt{2}) \) or \( (1/\sqrt{2},\: -1/\sqrt{2}) \),
it can be shown (as in ref. \cite{key-4}) that the three states \( (1/{\sqrt{2}})\left( \left| 00\right\rangle +\left| 11\right\rangle \right)  \),
\( (1/{\sqrt{2}})\left( \left| 00\right\rangle -\left| 11\right\rangle \right)  \)
and \( (1/{\sqrt{2}})\left( \left| 01\right\rangle +{\textrm{exp}}(i\phi )\left| 10\right\rangle \right)  \)
(or \( (1/{\sqrt{2}})\left( \left| 01\right\rangle -{\textrm{exp}}(i\phi )\left| 10\right\rangle \right)  \))
are locally indistinguishable with certainty if only a single copy
is provided. Similarly, it can be shown that the three states \( (1/{\sqrt{2}})\left( \left| 01\right\rangle +\left| 10\right\rangle \right)  \),
\( (1/{\sqrt{2}})\left( \left| 01\right\rangle -\left| 10\right\rangle \right)  \)
and \( (1/{\sqrt{2}})\left( \left| 00\right\rangle +{\textrm{exp}}(i\phi )\left| 11\right\rangle \right)  \)
(or \( (1/{\sqrt{2}})\left( \left| 00\right\rangle -{\textrm{exp}}(i\phi )\left| 11\right\rangle \right)  \))
are locally indistinguishable with certainty if only a single copy
is provided.}{\small \par}
\end{thebibliography}
\end{document}